\documentclass[12pt]{article}
\newcommand{\fr}{ \frac}

\begin{document}
\begin{center}
{\Large\bf Casimir Energy in a Conical Wedge and a Conical Cavity}
\end{center}
\vspace{5mm}
\begin{center}
H. Ahmedov and I. H. Duru
\end{center}

\noindent Feza G\"ursey Institute, P.O. Box 6, 81220,
\c{C}engelk\"{o}y, Istanbul, Turkey \footnote{E-mail :
hagi@gursey.gov.tr and duru@gursey.gov.tr}.

\vspace{5mm}
\begin{center}
{\bf Abstract}
\end{center}
Casimir energies  for a massless scalar field for a conical wedge
and a conical cavity are calculated. The  group generated by the
images is employed in deriving the Green function as well as the
wave functions and the energy spectrum.

 \vspace{2cm}\noindent {\bf
I. Introduction }
\\
\\
Image method is one of the best tools in the calculation of the
Casimir energies for the polygonal regions. One can use discrete
group generated by the reflections with respect to the boundaries
of the polygon under consideration to obtain the Green function,
wave functions and spectra of the system. The method is trivially
applicable to the cavities or the given regions of  space if the
following  conditions are fulfilled: Surfaces of the region are
planar; and, they are parallel or perpendicular to each other. The
discrete groups generated by reflections with respect to the
planar boundaries appears to be the direct product of the
translation $Z$ and the cyclic $C_2$ groups. The number of copies
of $Z$ and $C_2$ depend on the dimension of the rectangle. For the
parallel plate system and two dimension rectangle we have the
groups $C_2\times Z$ and $(C_2\times Z )^2$ respectively.  ( For a
Casimir calculation in rectangular region we refer to the
monograph \cite{CUBE} and references therein ).
\\
\\
Recently we tried to extend the image method to a set of
geometries with planar boundaries without rectangular angles
\footnote{ To our knowledge only previous example without
rectangular angles was the wedge problem \cite{WEDGE}. }. For a
class of cavities with triangular crossections, making use of the
tools of the group of reflections from the walls one can obtain
the Green function satisfying the Dirichlet boundary conditions,
then calculates the Casimir energy \cite{AHMED}. Point groups and
crystallographic point groups play important role in constructing
the Green functions and the wave functions if the region we
consider is the fundamental domain of the one of those groups. In
other words, if every point in the region represents different
orbits of the group, the satisfaction of the required wave
equation by the Green function is achieved \cite{AHMED},
\cite{AHMEDP}.
\\
\\
In this work the example we present is the application of the
image method to a geometry with non-planar boundary. The surface
of the cone we deal with is obtained from a geometry with planar
surfaces, by identifying two of these surfaces. In fact if there
is an element of the point group which maps one planar boundary
onto other one, we can obtain volumes with non planar boundaries.
The simple example we can think of is the  wedge in  $R^2$ with
the angle $\frac{2\pi}{N}$. By identifying two lines which are the
boundaries of the wedge we arrive at the conic surface with the
opening angle $2\arcsin \frac{1}{N}$. The corresponding point
group is the cyclic one $C_N$, that is generated by the element
$q$ which identifies two boundaries of the wedge. It is clear that
$q^N=1$.
\\
\\
In the next  section we first consider a conical wedge in $R^3$
with the opening angle $2\arcsin \frac{1}{3}$. The region  is the
fundamental domain of the tetrahedral group which has $C_3$ as
subgroup. We construct the Green function, then calculate   the
Casimir energy density for the massless scalar field.
\\
\\
 In section III making use of the group of the previous section we construct the
 wave function and  calculate the energy spectrum for a the conical wedge.
 \\
\\
Section IV  is devoted to the closed cone, i.e., conical cavity
for which we calculate the Casimir energy which is positive.
\\
\\
In the last section we briefly reviewed the known Casimir energy
results for  3-dimensional cavities.
\\
\\
The required group theoretical details are given in the appendix.
\\
\\
{\bf II. Green Function and Casimir Energy in a  Conical Wedge  }
\\
\\
Consider the following three planes in the first quadrant (
$x_1>0, \ x_2>0, \ x_3>0$ )  ( Figure 1)
\begin{equation}\label{planes}
P_1: \ x_1=x_2, \ \ \ P_2: \ x_2=0. \ \ \ P_3: \ x_2=x_3.
\end{equation}
We are interested in the Green function for the massless scalar
field satisfying  the  boundary conditions
\begin{equation}\label{B1}
K(x,y)|_{\vec{x}\in P_2}=0
\end{equation}
and
\begin{equation}\label{B2}
K(x,y)|_{\vec{x}\in P_1}=K(x,y)|_{g\vec{x}\in P_3}
\end{equation}
where
\begin{equation}
g = \left(
\begin{array}{ccc}
0 &  0 & 1 \\
1 & 0 & 0 \\
0 & 1 & 0
\end{array}
\right )
\end{equation}
is the rotation matrix around $\vec{n}=(1,1,1)$; i.e., the
intersection of $P_1$ and $P_2$, by angle $\frac{2\pi}{3}$. It
transforms $P_1$ into $P_3$. The region under consideration
\begin{equation}\label{domain}
 S=\{ x_1\geq x_2\geq 0, \ \  x_3\geq x_2\geq 0 \}
\end{equation}
is the fundamental domain $S=R^3/G$ of the group $G=T\times I$,
where $T$ is the tetrahedral group and $I$ is the inversion group
generated by $i=-1$ ( see Appendix ). Identifying planes $P_1$ and
$P_3$ in the manner described in (\ref{B2})  we arrive at the
space which is topologically equivalent to a cone with the plane
$P_2$ being the boundary  ( Figure 2 ). Bringing $\vec{x}_1$ onto
$\vec{x}_3$ axis requires a rotation by the angle $\frac{2\pi}{3}$
around $\vec{n}$. Thus the cone we obtained has an opening angle
$2\beta =2\arcsin \frac{1}{3}\cong 39^{\circ}$. Therefore the
problem in hand is equivalent to the study of the cone with
Dirichlet boundary condition.
\\
\\
The desired Green function is of the form
\begin{equation}\label{Green}
K(x,y) =\sum_{j=0}^{11} [ G( g_j x, y)-G( ig_j x, y)],
\end{equation}
where summation runs over the 12 elements of the tetrahedral
group. Here $G(x,y)$  is the free Green function for the massless
scalar field in the Minkowski space ( in $\hbar = c=1$ units )
\begin{equation}
G(x,y)=\fr{1}{4\pi^2} \fr{1}{\mid x-y\mid^2}.
\end{equation}
To check the boundary conditions, we observe that for any element
$g_j\in T$ there exist an element $g_a\in T$ such that $g_a=
g_jg$. Therefore
\begin{equation}
K(gx,y)=K(x,y)
\end{equation}
which implies that the boundary condition (\ref{B2}) is fulfilled.
To show the boundary condition (\ref{B1}) we first define the
reflection operators with respect to $P_2$
\begin{equation}
q = \left(
\begin{array}{ccc}
1 &  0 & 0  \\
0 & -1 & 0 \\
0 & 0 & 1
\end{array}
\right )
\end{equation}
For any element $g_j\in T$ there exist an element $g_b\in T$ such
that $ig_b= g_jq$. Therefore
\begin{equation}
K(qx,y)=-K(x,y)
\end{equation}
which implies that the boundary condition (\ref{B1}) is fulfilled.
\\
\\
To obtain the energy momentum tensor for the  conformally coupled
massless scalar field we employ the well known coincidence limit
formula  \cite{DAVIES}
\begin{equation}
T_{\mu\nu}=\lim_{x\rightarrow y}
[\frac{2}{3}\partial^{y}_{\mu}\partial^{x}_{\nu}-\frac{1}{6}(\partial^{x}_{\mu}\partial^{x}_{\nu}+
\partial^{y}_{\mu}\partial^{y}_{\nu}) -
\frac{\eta_{\mu\nu}}{6}\eta^{\sigma\rho}\partial^{y}_{\sigma}\partial^{x}_{\rho}+
\frac{\eta_{\mu\nu}}{24}\eta^{\sigma\rho}(\partial^{x}_{\sigma}\partial^{x}_{\rho}+\partial^{y}_{\sigma}\partial^{y}_{\rho})]
K(x,y)
\end{equation}
The energy density $T(x)=T_{00}$ is given by:
\begin{equation}\label{ED}
 T(x) =\frac{1}{12\pi^2}\sum_{j=1}^{11}[ T(g_j)-T(ig_j)]
\end{equation}
 where
\begin{equation}\label{T}
T(g)=(\frac{tr(g)-1}{\mid\vec{\eta}\mid^4}-2\frac{|((1+g)\vec{\eta}|^{2}}{\mid\vec{\eta}\mid^6})
\end{equation}
 and
\begin{equation}
  \vec{\eta} = (1-g)\vec{x}
\end{equation}
with $g$ standing for $g_j$ and $ig_j$.
\\
\\
Using the invariance of $T(g)$ under   $g\rightarrow g^{-1}$ we
have
\begin{eqnarray}\label{INV}
T(g_{4}) &=& T(g_{5}) \nonumber \\
T(g_{6}) &=& T(g_{7}) \\
T(g_{8}) &=& T(g_{9}) \nonumber \\
T(g_{10}) &=& T(g_{11}) \nonumber
\end{eqnarray}
The same equalities are  true for elements $ig_j$, with $j$
running the same values as (\ref{INV}).
\\
\\
$\bullet$ For the rotation $g_{6}$  by  angle $\fr{2\pi}{3}$
around the line passing trough the origin and the point $(1,-1,1)$
we have
\begin{equation}\label{2}
T(g_{6})= -\fr{3}{((x_1+x_2)^2+(x_2+x_3)^2+(x_3-x_1)^2)^2}
\end{equation}
\begin{equation}\label{20}
T(ig_{6})=-\fr{6|\vec{x}|^2+2(x_1x_3-x_1x_2-x_2x_3}{((x_1-x_2)^2+(x_2-x_3)^2+(x_3+x_1)^2)^3}.
\end{equation}
Since $g_{8}$ and $g_{10}$ are rotation matrices by the same angle
around the axis passing trough the origin and the points
$(-1,1,1)$ and $(1,1,-1)$ we conclude that $T(g_{8})$, $T(ig_{8})$
and $T(g_{10})$, $T(ig_{10})$  are given by (\ref{2}) and
(\ref{20}) with the cyclic replacements of coordinates
$(x_1,x_2,x_3)\rightarrow (x_3,x_1,x_2)$ and
$(x_1,x_2,x_3)\rightarrow (x_2,x_3,x_1)$ respectively.
\\
\\
$\bullet$ We also have
\begin{eqnarray}
T(g_{4}) = -\fr{3}{((x_1-x_2)^2+(x_2-x_3)^2+(x_3-x_1)^2)^2}, \\
T(ig_{4})= -\fr{6\mid \vec{x}\mid^2 -2(x_1x_2+x_1x_3+x_2x_3)}
{((x_1+x_2)^2+(x_1+x_3)^2+(x_2+x_3)^2)^3}.
\end{eqnarray}
\\
\\
$\bullet$ For elements satisfying the condition $g^2=1$ the second
expression in (\ref{T}) vanishes. These are the elements $g_j$,
$j=1,2,3$ .  Since $tr (ig_j)=1$  we have $T(ig_j)=0$. Nonzero
ones are then
\begin{equation}
 T(i)=-\frac{1}{4|\vec{x}|^4}.
\end{equation}
and
\begin{equation}\label{end}
 T(g_1)=-\fr{1}{8(x_2^2+x_3^2)^2}.
\end{equation}
Remaining two terms $T(g_{2})$, $T(g_{3})$ are   obtained from the
above  equation  by the cyclic replacements of coordinates.
\\
\\
Inserting the results from (\ref{2}) to (\ref{end}) into
(\ref{ED}) we arrive at  the energy density
\begin{eqnarray}\label{final}
 T(x) = \frac{1}{12\pi^2} [ T(g_{4})- T(ig_{4})-T(i)]  \ \ \ \ \
 \ \ \ \ \ \ \ \ \ \ \ \ \ \ \ \ \ \ \ \ \ \ \ \ \ \ \ \ \ \ \ \ \ \ \ \ \\
+\frac{1}{12\pi^2}[ \frac{1}{8}T(g_1)+2T(g_{6})  - 2T(ig_{6})+
c.p. ]\nonumber .
\end{eqnarray}
Here $c.p.$ stands for cyclic permutations of coordinates.
\\
\\
{\bf III. Wave Function and the Spectrum in the  Conical  Wedge}
\\
\\
In this section we briefly present the derivation of the wave
function and the spectrum. We write the wave function in a similar
fashion as the Green function as
\begin{equation}\label{wave0}
\Psi_{\vec{p}} (\vec{x})= \Omega  \sum_{j=0}^{11}[
e^{i(\vec{p},g_j\vec{x})}-e^{i(\vec{p},ig_j\vec{x})}]
\end{equation}
or
\begin{equation}\label{wave}
\Psi_{\vec{p}} (\vec{x})= -8i\Omega [\sin p_1x_1\sin p_2x_2\sin
p_3x_3 + \ c. p. ]
\end{equation}
with $\Omega$ being  the normalization. To obtain the spectrum we
first observe that
\begin{equation}
\Psi_{g_j\vec{p}} (\vec{x})=\Psi_{\vec{p}} (\vec{x}) \ \ \
\Psi_{i\vec{p}} (\vec{x})=-\Psi_{\vec{p}} (\vec{x})
\end{equation}
which imply that the momentum   $\vec{p}$ takes its values in the
quotient space $R^3/G$ which is exactly the region $S$ of
(\ref{domain}). In other words, the geometry in the momentum space
is the replica of the geometry in the configuration space, the
fact that can be used in deriving the spectra. Since
\begin{equation}\label{W1}
\Psi_{\vec{p}} (\vec{x})|_{\vec{p}\in P_2}=0
\end{equation}
we have to drop the values of $\vec{p}$ on the plane $P_2$
($p_2=0$ ). The condition
\begin{equation}\label{W2}
\Psi_{\vec{p}} (\vec{x})|_{\vec{p}\in P_1}=\Psi_{\vec{p}}
(\vec{x})|_{g\vec{p}\in P_3}
\end{equation}
implies that  momenta  on the plane $P_1$ ( $p_1=p_2$ ) are
equivalent to the ones on the plane $P_3$ ( $p_2=p_3$ ). It is
sufficient to take into account momenta on one of these planes (
take for example values on $P_1$ ) and drop the ones on the other.
Therefore the spectrum in the conical wedge  is
\begin{equation}\label{spectra}
\Sigma_0 =S\setminus (A\bigcup B)= \{ p_1\geq p_2> 0, \ \ p_3>
p_2> 0 \}
\end{equation}
where $A$ and $B$ are the boundaries of $S$ corresponding to $P_2$
and $P_3$ planes:
\begin{equation}
 A=\{ p_1\geq  0, \ \ p_2=0 \ \  p_3\geq 0 \},
\end{equation}
\begin{equation}
 B=\{ p_1\geq p_2\geq 0, \ \ p_3=p_2\geq 0 \}.
\end{equation}
\\
\\
{\bf IV. Conical Cavity}
\\
\\
In addition to the planes of (\ref{planes}) consider two
additional ones
\begin{equation}
 P_4: \ x_1=a, \ \ \ \  P_5: \ x_3=a.
\end{equation}
The region we consider is then given by
\begin{equation}
 H=\{ a\geq x_1\geq x_2\geq 0, \ \  a\geq x_3\geq x_2\geq 0 \}.
\end{equation}
Boundary conditions (\ref{W1}), (\ref{W2}) and new Dirichlet
conditions on $P_4$ and $P_5$
\begin{equation}\label{W3}
\Psi_{\vec{p}} (\vec{x})|_{\vec{p}\in P_4}=\Psi_{\vec{p}}
(\vec{x})|_{\vec{p}\in P_5}=0
\end{equation}
imply that the geometry in hand is a conical cavity with the
opening angle $\beta = \arcsin \frac{1}{3}$ and height
$h=a\cos\beta = \frac{2\sqrt{2}}{3}a$. The Dirichlet boundary
conditions (\ref{W3}) quantize the momenta:
\begin{equation}
p_1=\frac{\pi}{a} n, \ \ p_2=\frac{\pi}{a} m, \ \
p_3=\frac{\pi}{a} k
\end{equation}
where $n$, $m$ and $k$ are integers which subject to the condition
(\ref{spectra}) ( with $\vec{n}=(n,m,k)$ )
\begin{equation}
\Sigma = \Sigma_0 \bigcap Z^3= \{ \vec{p}=\frac{\pi}{a}\vec{n}: \
\ n\geq m> 0, \ \ k> m> 0 \}.
\end{equation}
\\
\\
The Green function can be written as
\begin{equation}
G(x,y)=\sum_{m=1}^\infty\sum_{k=m+1}^{\infty}\sum_{n=m}^{\infty}\fr{e^{i\pi\mid
\vec{p}\mid (x_0-y_0)}} {2\mid\vec{p}\mid}\Psi_{\vec{p}} (\vec{x})
\Psi_{\vec{p}} (\vec{y}).
\end{equation}
The energy density we obtain is
\begin{equation}
 T(x)=\fr{\pi}{2a}\sum_{m=1}^\infty\sum_{k=m+1}^{\infty}\sum_{n=m}^{\infty}
\sqrt{n^2+m^2+k^2}\mid \Psi_{\vec{p}} (\vec{x})\mid^2.
\end{equation}
After integration $\int_0^a dx_2\int_{x_2}^a dx_3\int_{x_2}^a
dx_1$  over the conical cavity $H$ we get the total energy
 \begin{equation}\label{Eq1}
E=\frac{\pi}{2a}\sum_{\vec{n}\in \Sigma}|\vec{n}|=
\frac{\pi}{48a}\sum_{\vec{n}\in \Sigma}\sum_{g\in G}|g\vec{n}|=
\frac{\pi}{48a} \sum_{\vec{n}\in \bigcup g\Sigma}|\vec{n}|.
\end{equation}
 Since $\Sigma = S\backslash (A\bigcup B)) \bigcap Z^3$ and   $S=R^3/G$ we have
\begin{equation}\label{Eq2}
 \bigcup_{g\in G} g\Sigma= Z^3\setminus C, \
\end{equation}
where
\begin{equation}
 C = (\bigcup_{g\in G} g (A\bigcup B))\bigcap Z^3
\end{equation}
is the union of six planes in $Z^3$: $m=0$, $k=m$ and other four
planes are obtained by cyclic permutations of $n, m$ and $k$: .
\begin{equation}\label{Eq3}
 \sum_{\vec{n}\in C}|\vec{n}|=3\sum_{n,m\in Z}\sqrt{n^2+m^2}+3\sum_{n,m\in Z}\sqrt{2n^2+m^2}
 \end{equation}
(\ref{Eq1}) and (\ref{Eq2}) imply
\begin{equation}
 E=\frac{\pi}{48a}( \sum_{\vec{n}\in Z^3}|\vec{n}| -
 \sum_{\vec{n}\in C}|\vec{n}| )
 \end{equation}
 or
\begin{equation}\label{CASIMIR}
E= \frac{1}{3}E _1-\frac{1}{2}E_2-\frac{2+\sqrt{2}}{4}E_3 \simeq
\frac{0,085}{a}.
\end{equation}
 Here $E_1$, $E_2$ and  $E_3$  are the Casimir energies for the cube with sides
 $a$,  for the  rectangle with sides $a$ and  $\frac{a}{\sqrt{2}}$; and, for the one dimensional system of length
 $a$   respectively \cite{TRUN}:
\begin{eqnarray}
E_1&=&\frac{\pi}{2a}\sum_{n,m,k=1}^\infty \sqrt{n^2+m^2+k^2}\simeq   -\frac{0,015}{a}, \\
E_2&=&\frac{\pi}{2a}\sum_{n,m=1}^\infty \sqrt{n^2+2m^2} \simeq
\frac{0,045}{a},     \\
E_3&=&\frac{\pi}{2a}\sum_{n=1}^\infty n  \simeq -\frac{0,131}{a}.
\end{eqnarray}
\\
\\
{\bf Discussion}
\\
\\
Examples of the Casimir energy calculations for 3-dimensional
cavities are not plenty. The cubical \cite{CUBE, TRUN} and the
spherical \cite{BALL} cavities have been studied rather
extensively. In this work we have studied a conical cavity with a
particular opening angle. Similar restriction  is true for the
recently studied pyramidal geometry for which the angles at the
vertices are not arbitrary \cite{AHMEDP}.  We do not have results
in hand for the last two kind of geometries with arbitrary angles.
Nevertheless it may be of interest to briefly review the known
Casimir energy calculations ( for the massless scalar fields ) for
the above mentioned geometries. To have better idea on the
magnitudes we consider the cavities of comparable sizes:
\\
\\
For cube of sides $2b$, we have negative value for the energy :
$E_{cub}\simeq -\frac{0,007}{b}$ \cite{CUBE, TRUN}.
\\
\\
For spherical cavity of radius $b$ the Casimir energy is positive
$E_{sph}\simeq \frac{0,046}{b}$ \cite{BALL}.
\\
\\
Coming to the recently studied pyramidal cavity, let us first
describe its position with respect to the above cube.  One of the
four vertices is located at the center of the  cube, the remaining
three at the center of a surface, and at the closest vertex and at
the middle of the closest edge to this vertex \cite{AHMEDP}. The
volume of this cavity is $\frac{b^3}{6}$ and the Casimir energy is
again positive, $E_{pyr}\simeq \frac{0,069}{b}$.
\\
\\
For the present conical cavity of opening angle
$2\beta=2\arcsin\frac{1}{3}$ and height b ( i.e. $b=
\frac{2\sqrt{2}}{3}a$ ) which has the positive energy of
(\ref{CASIMIR}) is $E_{con}\simeq \frac{0,08}{b}$.
\\
\\
To see the dependence of the magnitudes of the positive Casimir
energies on the "shapes" of the cavities, let us compare the last
three results for the "equal" volumes. If we consider the
spherical, pyramidal and conical cavities with equal volumes we
have the following ratios
\begin{eqnarray}
E_{con}\simeq 0,54 E_{sph} \\
E_{pyr}\simeq 0,51 E_{sph}.
\end{eqnarray}
It is not surprising that the conical and pyramidal geometries are
very close to each other.
\\
\\
{\bf Acknowledgments}: We thank Turkish Academy of Sciences  (
TUBA ) for its support, and  D. Duru for the figures.
\\
\\
\setcounter{equation}{0}
\def\theequation{A.\arabic{equation}}

\vspace{1cm} \noindent {\bf Appendix A}
\\
\\
Tetrahedral  group $T$ is the group of transformations which
transforms cube into itself. The order of this group is 12. We
denote the identity element by $g_0$. $g_1$, $g_2$ and $g_3$ are
rotations on $\pi$ around x, y and z-axis:
\begin{equation}
g_1 = \left(
\begin{array}{ccc}
1 &  0 & 0  \\
0 & -1 & 0 \\
0 & 0 & -1
\end{array}
\right ),
g_2 = \left(
\begin{array}{ccc}
-1 & 0 & 0 \\
0 & 1 & 0  \\
0 & 0  & -1
\end{array}
\right ), \\
g_3 = \left(
\begin{array}{ccc}
-1 & 0 & 0 \\
0  & -1 & 0 \\
0 & 0 & 1
\end{array}
\right )
\end{equation}
$g_{4}$ and  $g_{5}$  are rotations by $\fr{2\pi}{3}$ and
$\fr{4\pi}{3}$ around the axis passing trough the origin and the
point $(1,1,1)$:
\begin{equation}
g_{4} = \left(
\begin{array}{ccc}
0 &  0 & 1  \\
1 &  0 & 0 \\
0 & 1 & 0
\end{array}
\right ),
 g_{5} = \left(
\begin{array}{ccc}
0 & 1 & 0  \\
0 & 0 & 1 \\
1 & 0 & 0
\end{array}
\right ),
\end{equation}
$g_{6}$ and  $g_{7}$  are rotations by $\fr{2\pi}{3}$ and
$\fr{4\pi}{3}$ around the axis passing trough the origin and the
point $(1,-1,1)$:
\begin{equation}
g_{6} = \left(
\begin{array}{ccc}
0 &  0 & 1  \\
-1 &  0 & 0 \\
0 & -1 & 0
\end{array}
\right ), g_{7} = \left(
\begin{array}{ccc}
0 & -1 & 0  \\
0 &  0 & -1 \\
1 & 0 & 0
\end{array}
\right )
\end{equation}
$g_{8}$ and  $g_{9}$  are rotations by $\fr{2\pi}{3}$ and
$\fr{4\pi}{3}$ around the axis passing trough the origin and the
point $(-1,1,1)$:
\begin{equation}
g_{8} = \left(
\begin{array}{ccc}
0 &  0 & -1  \\
-1 &  0 & 0 \\
0 & 1 & 0
\end{array}
\right ), \\
g_{9} = \left(
\begin{array}{ccc}
0 &  -1 & 0  \\
0 &  0 & 1 \\
-1 & 0 & 0
\end{array}
\right )
\end{equation}
$g_{10}$ and  $g_{11}$  are rotations by $\fr{2\pi}{3}$ and
$\fr{4\pi}{3}$ around the axis passing trough the origin and the
point $(1,1,-1)$:
\begin{equation}
g_{10}= \left(
\begin{array}{ccc}
0 &  0 & -1  \\
1 &  0 & 0 \\
0 & -1 & 0
\end{array}
\right ), g_{11} = \left(
\begin{array}{ccc}
0 &  1 & 0  \\
0 &  0 & -1 \\
-1 & 0 & 0
\end{array}
\right ).
\end{equation}
\\
\\
Let $G$ be a point group acting in the Euclidean  space  $R^3$. A
closed subset $S$ of $R^3$ is called a fundamental domain of $G$
 if $R^3$ is the union of conjugates of $S$, i.e.,
\begin{equation}
 R^3=\bigcup_{g\in G} gS
\end{equation}
and the intersection of any two conjugates has no interior.
\\
\\
The fundamental domain of the group generated by the reflections
$ig_1$, $ig_2$ and $ig_3$ with respect to $x_1=0$, $x_2=0$ and
$x_3=0$ planes is the first quadrant in $R^3$. This is group of
order $8$ and divide $R^3$ into $8$ equal parts. If one adds to
this group the element  $g_4$ we arrive at the group of order $24$
which is the direct product of the tetrahedral group $T$ and
inversion one $I$ generated by $i$. Rotation $g_4$ is three fold
rotation. It divides the first quadrant into three equal parts.
Therefore the fundamental domain for $T\times I$ is the region in
the first quadrant between three planes (\ref{planes}).
\\
For more details concerning finite groups we refer to \cite{Jans}.

\end{document}